\documentclass[preprint]{revtex4-2}
\usepackage[ansinew]{inputenc}
\usepackage[T1]{fontenc}
\usepackage{float}
\usepackage{amsmath}
\usepackage{ulem}
\usepackage{mathptmx}
\usepackage{mathrsfs}
\usepackage[english]{babel}
\usepackage{graphicx}
\usepackage{hyperref}
\usepackage[usenames,dvipsnames]{xcolor}
\usepackage{natbib}
\usepackage{makecell}
\usepackage[per-mode=reciprocal, range-units = single, range-phrase=\ -\ , exponent-product = \cdot , separate-uncertainty = true, multi-part-units=single , quotient-mode = fraction]{siunitx}

\usepackage{blindtext}
\usepackage{physics}

\begin{document}

\preprint{AIP/123-QED}

\title{Low-temperature thermal conductivity of the substrate material YAlO$_3$ and its unconventional sister compound YbAlO$_3$}

\author{Parisa Mokhtari}
\altaffiliation[Now at ]{Department of Applied Physics and Quantum-Phase Electronics Center, The University of Tokyo, Bunkyo-ku, Tokyo 113-8656, Japan.}
\affiliation{Department of Physics, Technical University of Munich, 85748 Garching, Germany}
\affiliation{Max Planck Institute for Chemical Physics of Solids, 01187 Dresden, Germany}
\affiliation{Faculty of Physics, Technische Universit\"at Dresden, 01062 Dresden, Germany}

\author{Ulrike Stockert}%
 \email{ulrike.stockert@tu-dresden.de}
\affiliation{Faculty of Physics, Technische Universit\"at Dresden, 01062 Dresden, Germany}

\author{Stanislav E. Nikitin}
\affiliation{PSI Center for Neutron and Muon Sciences, 5232 Villigen PSI, Switzerland}

\author{Leonid Vasylechko}
\affiliation{Lviv Polytechnic National University, Lviv 79013, Ukraine}

\author{Manuel Brando}
\affiliation{Max Planck Institute for Chemical Physics of Solids, 01187 Dresden, Germany}

\author{Elena Hassinger}
\affiliation{Faculty of Physics, Technische Universit\"at Dresden, 01062 Dresden, Germany}
\affiliation{Max Planck Institute for Chemical Physics of Solids, 01187 Dresden, Germany}

\date{\today}% It is always \today, today,
             %  but any date may be explicitly specified

\date{\today}% It is always \today, today,
             %  but any date may be explicitly specified

\begin{abstract}
We present thermal conductivity data on single crystals of YAlO$_3$ and YbAlO$_3$ for temperatures between 2~K and 300~K and the heat current along $b$ and $c$. Both materials are very good thermal conductors in the investigated temperature range. The thermal conductivity in these electrical insulators is due to phonons. The effect of Y-Yb exchange is found to be rather small despite the considerable difference in density and average atomic mass. For YAlO$_3$ we find a moderate thermal conductivity anisotropy with weak temperature dependence and a ratio of $c$ to $b$ direction between at most 1 and 2.2. It is discussed with regard to the velocities of sound and relevant scattering processes. For YbAlO$_3$ the small crystal size limits the precision of absolute thermal conductivity values and does not allow drawing conclusions on the anisotropy. Our results on YAlO$_3$ confirm that the material is suitable for applications requiring a good thermal conductivity at temperatures down to liquid helium, such as lasers, substrates, and detectors. 
\end{abstract}

\maketitle

%xxxxxxxxxxxxxxxxxxxxxxxxxxxxxxxxxxxxxxxx

\section{Introduction}

YAlO$_3$ is widely used in applications, both as a pure compound and when doped with different elements~\cite{Vasylechko-2009}. Single crystalline YAlO$_3$, in literature also referred to as YAP (yttrium aluminum perovskite) is a popular substrate material for growing thin films, for example of high-temperature superconductors~\cite{Robles-2003}, heavy-fermion materials~\cite{Uchida-2017}, and doped perovskites~\cite{Dewo-2021}. It has been also suggested as a window material in photodetectors~\cite{Ambrosio-2000}.
Cr-doped YAlO$_3$, where Cr replaces part of the Al, has been proposed as a temperature sensor material~\cite{Uchiyama-2003}, while Mn:YAP is discussed as a candidate for data storage and holographic recording~\cite{Noginov-1998, Loutts-1998}. The largest potential for applications is accomplished by incorporation of rare-earth (RE) ions. RE:YAlO$_3$, where part of the yttrium is replaced by elements such as Pr, Nd, Gd, and Yb, are important laser and scintillator materials with possible use in medicine~\cite{Vittori-1999, Yamamoto-2022}, astrophysics~\cite{Costa-1995, Song-2023}, and synchrotron technology~\cite{Kishimoto-2003}. 
A good thermal conductivity is either mandatory or at least advantageous for most of these applications for instance to facilitate energy release in laser materials or thermal coupling of thin films via substrates to the bath. In fact, a good thermal conductivity is expected for high-quality single crystals of YAlO$_3$. Experimentally, the thermal conductivity of YAlO$_3$ has been measured in the temperature range between 80 K and 300 K \cite{Aggarwal-2005} and between 300 K and 870 K~\cite{Zhang-2023}. However, low-$T$ data are missing so far and cannot be extrapolated or estimated from the high-temperature values: The thermal conductivity of insulating single crystals does not change monotonously with temperature but goes through a maximum typically around 10-30 K before approaching zero in the zero-temperature limit. This characteristic $T$-dependence arises from cumulative phonon excitations upon heating and diverse scattering processes dominating in different $T$ ranges. It depends on sample size and quality~\cite{Berman-1976}.

The above-mentioned chemical substitution also has strong influence on the thermal conductivity. Partial substitution of one element by another usually lowers the thermal conductivity via the introduced disorder. Indeed, for Yb:YAlO$_3$ a significant lowering of the thermal conductivity has been confirmed experimentally~\cite{Aggarwal-2005, Petit-2010, Song-2020}.
Besides disorder, substitution has also another effect: The difference in atomic mass between the two involved elements changes the density and the phonon spectrum of the material. This effect may be quite large, e.g., exchanging Y with an atomic mass $m_{\mathrm{a}} = 88.9$~u by Yb with $m_{\mathrm{a}} = 173.0$ u increases the density from YAlO$_3$ to YbAlO$_3$ by a factor of 1.5. The Debye temperature $\Theta_\mathrm{D}$, which is related to the phonon spectrum, is also expected to change, however, literature values are not consistent~\cite{Aggarwal-2005, Zhydachevskii-2006, Senyshyn-2013}. Large atomic masses and low $\Theta_\mathrm{D}$ facilitate low $\kappa$ values at high $T$ (strictly at $T > \Theta_\mathrm{D}$)~\cite{Slack-1973}. In order to disentangle the effects of disorder, effective mass, and $\Theta_\mathrm{D}$, thermal conductivity data for YbAlO$_3$ are needed. 
At very high $T \gg \Theta_\mathrm{D}$ theoretical calculations of the structural and elastic properties  predict that the minimum thermal conductivity of YbAlO$_3$, $\kappa_{\mathrm{min}} = 1.15$~W/Km \cite{Xiang-2015}, is significantly lower than the one for YAlO$_3$ of $\kappa_{\mathrm{min}} = 1.61$~W/Km  \cite{zhan_theoretical_2012}. 
The only available experimental $\kappa$ values for YbAlO$_3$ are in the milli-Kelvin range in a regime with highly unconventional magnetic phases \cite{Mokhtari-2025}.

In fact, the low-$T$ phase diagram of YbAlO$_3$ has been studied intensively over the past few years, leading to the discovery of a Luttinger-liquid regime \cite{Wu-2019}, multiple antiferromagnetic phases~\cite{Wu-2019b} and several plateau states upon applying magnetic fields~\cite{Mokhtari-2025}. This complex low-$T$ behavior is also reflected in the thermal conductivity. A thorough understanding of $\kappa$ requires an evaluation of the contribution from phonons, either from an extrapolation from high $T$ data above the exotic region or from a suitable reference compound. These considerations stimulated our interest in the thermal conductivity of YAlO$_3$ and YbAlO$_3$ also from a fundamental point of view and beyond its relevance for applications.

In this manuscript we %partially fill the above-mentioned gaps of the existing experimental data. We
present the thermal conductivity of YAlO$_3$ between 2 K and 300 K thereby extending the available data range down to liquid He temperatures. We fit the data to the Callaway model and discuss the relevant scattering processes. In addition we show thermal conductivity data in the temperature window from 50 K to 300 K on YbAlO$_3$ for which the change in average atomic mass and also density compared to YAlO$_3$ is very large and only exceeded marginally by the one of LuAlO$_3$. Finally, we discuss the differences between the Y and Yb compounds as well as implications for applications. We also comment on the relevance of our results for understanding the low-$T$ thermal conductivity of YbAlO$_3$.

\section{Experimental Techniques}
Oriented single crystals of YAlO$_3$ grown by the Czochralski technique were bought from CRYSTAL GmbH, Berlin (lattice constants: $a = 5.18027$ \AA, $b = 5.32951$ \AA, and $c = 7.37059$ \AA). A single crystal of YbAlO$_3$ (lattice constants: $a = 5.126$ \AA, $b = 5.331$ \AA, and $c = 7.313$ \AA) was grown by a Czochralski technique as described elsewhere~\cite{buryy_thermal_2010}. The orientation of the YbAlO$_3$ crystal was initially determined by Laue diffraction and subsequently confirmed by magnetization measurements at low $T$ making use of the large angular dependence of the magnetic properties of the material~\cite{Wu-2019}. The oriented crystals were cut into cuboid bars, with dimensions given in Table \ref{tab:reference_samples}.

It needs to be noted that the orientation convention for YAlO$_3$ and related materials is not uniform in literature~\cite{Diehl-1975}, which becomes important when comparing results for different crystallographic directions. Throughout this manuscript we use the Pbnm setting of space group N 62, where $c > b \geq a$. 

The thermal conductivity $\kappa$ was measured using a standard steady-state method with a two-thermometer-one-heater configuration in the $T$ range of 2 K - 300 K in a commercially available Quantum Design (QD) Physical Property Measurement System (PPMS). %\textcolor{red}{
Two calibrated Cernox chip resistors sitting on gold-plated copper shoes were used as thermometers to determine the average sample temperature and the temperature difference along the sample.%} 
The heat current $j_Q$ was applied $\parallel \, b$ ($\kappa_b$) and $\parallel \, c$ ($\kappa_c$). To ensure thermal coupling of the contacts to the electrically insulating samples, four micro-contacts consisting of a 10~nm titanium layer covered by a 150 nm gold layer were sputtered onto each sample. One end of the samples was clamped at the cold bath. The contacts to the thermometers and the heater were accomplished via gold-plated copper bars glued to the %\textcolor{red}{
sample 
%} 
contact pads with silver paint %\textcolor{red}{
and clamped to the thermometer shoes by screws. %} 
The values for the mean distance $l_{\Delta T}$ between the cold and hot probes are given in Table \ref{tab:reference_samples}. The finite contact widths $\delta l_{\Delta T}$ of about 0.2 mm and the measurement accuracy of the sample dimensions of about $\pm 0.01$ mm sum up to a considerable uncertainty of the geometry factor and in consequence of the absolute values of $\kappa$, upper limits being about 11 \% for YAlO$_3$ and 21 \% for YbAlO$_3$. This systematic error does not affect the shape of $\kappa(T)$, but enters only as a prefactor changing the absolute $\kappa$ values. However, a discussion of differences or the anisotropy of $\kappa$ is not reasonable below these thresholds. 

\begin{table}
\caption{\label{tab:reference_samples}%
Dimensions and contact geometries of the measured samples.
    $l_a$, $l_b$, and $l_c$ are the sample lengths along the $a$-, $b$-, and $c$-axis, respectively. 
    $l_{\Delta T}$ is the mean distance between the cold and hot probes.}
\begin{ruledtabular}
\begin{tabular}{cccc}
     Sample & ($l_a \times l_b \times l_c$) (mm$^3$) &  $\Vec{j}_{Q}$ & $l_{\Delta T}$ (mm) \\
    \colrule
    YAlO$_3$ \#b & $0.5 \times 5 \times 0.5$ & $\parallel b$ & 2.7 \\
    YAlO$_3$ \#c &  $0.5 \times 0.5 \times 5$ & $\parallel c$ & 2.7 \\
    YbAlO$_3$ \#b &  $0.45 \times 2 \times 0.45$ & $\parallel b$ & 1.2 \\
    YbAlO$_3$ \#c &  $0.5 \times 0.5 \times 1.95$ & $\parallel c$ & 1.15 \\
\end{tabular}
\end{ruledtabular}
\end{table}

In order to avoid excessive heat loss from thermal radiation during measurements, the rise of the average sample temperature $T_\mathrm{av}$ was limited to about 3 \% compared to the bath temperature. Depending on the thermal coupling between sample and bath and the sample geometry, the resultant temperature difference between warm and cold probe for YAlO$_3$ was typically within 1 \% and 3 \% of $T_\mathrm{av}$. 

%\textcolor{red}{
The radiation loss was estimated from the sample temperatures at the hot and cold contacts, the bath temperature, the sample surface area, and a rough value for the emissivity of the materials of 0.5. Around 250~K the calculated radiation loss exceeds 1~\% of the heater power  and reaches at most 2.5~\% close to room temperature. However, we know from experiments, where we compared thermal conductivity data obtained with different methods, that radiation losses are underestimated by our correction procedure and may get important already above about 200~K~\cite{Sun-2025}. Therefore, we included a term accounting for radiation losses in our model.%}

For YbAlO$_3$ the very high thermal conductivity in combination with the small sample size led to extremely small or even unresolvable temperature gradients below about 50 K. Despite our efforts to improve the thermal coupling by using gold-plated contacts, we were not able to obtain reproducible data for this material at lower $T$. Therefore, we restrict our presentation and discussion in this case to temperatures above 50 K, i.e. covering the regime of liquid nitrogen, which is still an important temperature range for applications.

\section{Results and Discussion}
\subsection{Heat transport in YAlO$_3$}

\begin{figure}[t!] %{wrapfigure}{r}{0.55\textwidth}
\centering 
 \includegraphics[width=0.6\columnwidth]{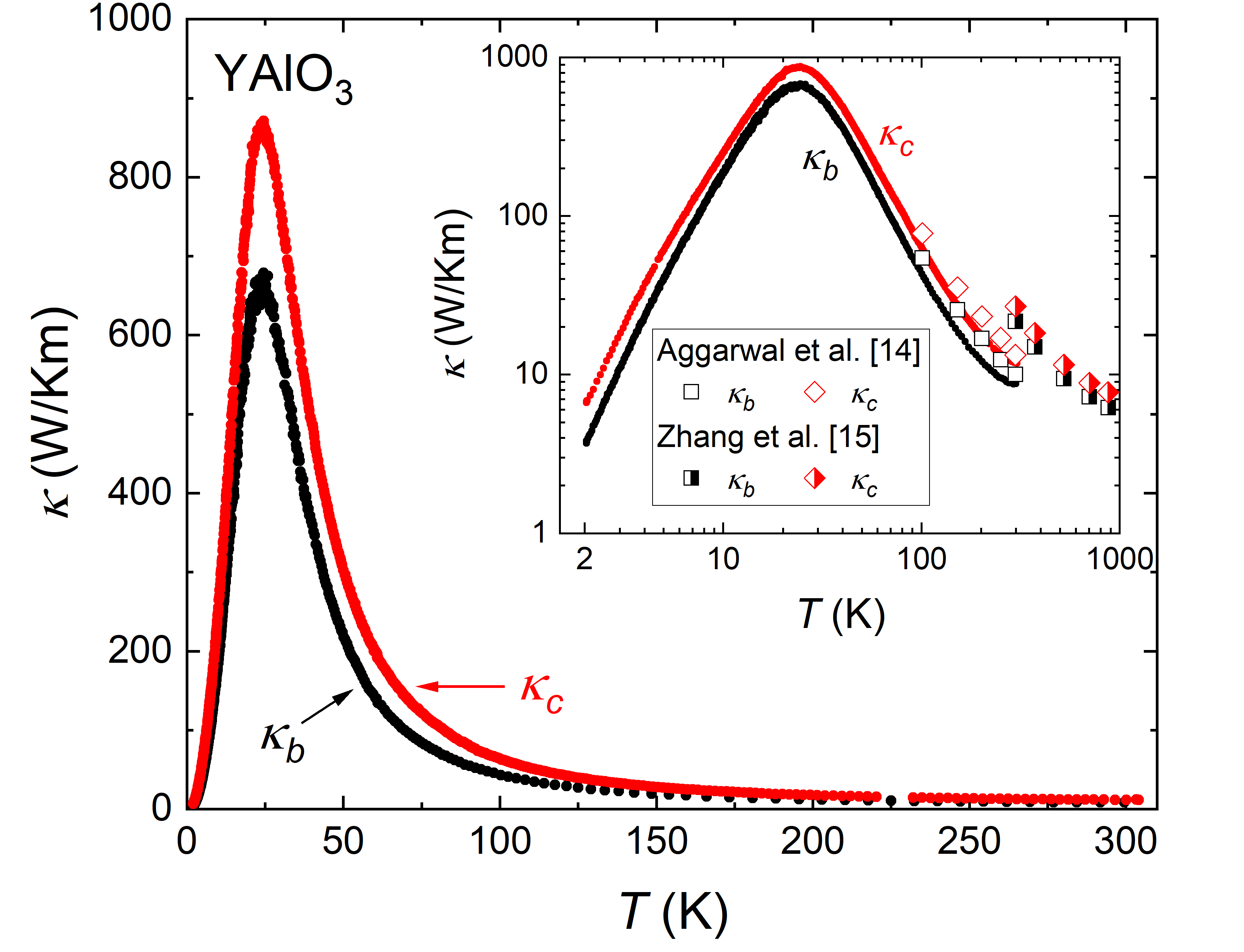}
\caption{Temperature dependence of the thermal conductivity \(\kappa\) of YAlO$_3$ 
along the \(b\)-axis (\(\kappa_b\)) and the \(c\)-axis (\(\kappa_c\)) plotted in black and red, respectively. The inset presents the same data on a double-logarithmic scale in comparison to literature data from Aggarwal \textit{et al.} \cite{Aggarwal-2005} and Zhang \textit{et al.}~\cite{Zhang-2023}. %The blue line demonstrates the approximate scaling between $\kappa_b$ and $\kappa_c$.
}
\label{YAlO_kappa}
\end{figure}

Fig. \ref{YAlO_kappa} shows the thermal conductivity of YAlO$_3$ measured for the heat current $j_Q$ along $b$ ($\kappa_b$) and along $c$ ($\kappa_c$). The inset depicts the same data in a double logarithmic presentation in comparison to results from literature explained in detail below. The results for YAlO$_3$ can be directly compared to the equivalent data of YbAlO$_3$ shown in Fig. \ref{YbAlO_kappa}, that will be presented in the next section. 

The thermal conductivities $\kappa_b$ and $\kappa_c$ of YAlO$_3$ have very similar temperature dependencies, although with slightly different absolute values. They strongly increase with increasing temperature, go through huge maxima at about 25 K and decrease rapidly towards higher $T$. The values at the maximum are about 20 times larger than those in SrTiO$_3$, another popular substrate material with perovskite structure~\cite{Behnia-2018}. They also exceed those measured on isostructural NdGaO$_3$, and the related NdAlO$_3$ and LaAlO$_3$ by a factor of about 5~\cite{Morelli-1992, Schnelle-2001}. However, they are still well below those reached by other common substrate materials with different structure as silicon and sapphire, reaching 4 000 W/Km and more than 10 000 W/Km, respectively~\cite{McCurdy-1970, Slack-1962}.

The observed type of $T$ dependence is typical for clean, nonmetallic, and nonmagnetic single crystals: Thermal transport is purely phononic and limited by boundaries and defects at lowest $T$. At low $T$ $\kappa$ increases due to cumulative excitation of phonons. The maximum at mid $T$ arises due to the onset of umklapp (U) scattering of phonons with large momentum leading to a decreasing mean free path up to room temperature.

The thermal conductivity is smaller for $j_Q \parallel b$. It can be roughly scaled to the one for $j_Q \parallel c$ by a factor of 1.35. The ratio $\kappa_c/\kappa_b$ is plotted in Fig.~\ref{anisotropy} and discussed in more detail below. % with only small deviations below 10 K. This scaling is demonstrated by the line in the inset of Fig.~\ref{YAlO_kappa}. 
Our observation of an only moderate anisotropy is in accordance with literature results~\cite{Aggarwal-2005,Zhang-2023}. However, the relative size differs: While $\kappa_c > \kappa_b$ in our study in line with Ref.~\onlinecite{Zhang-2023}, an opposite anisotropy with $\kappa_b > \kappa_c$ has been %\textcolor{red}{
claimed%} 
in Ref.~\onlinecite{Aggarwal-2005}. This latter proportion has been challenged by A.~Hofmeister~\cite{Hofmeister-2010}: Based on thermal diffusivity measurements on YAlO$_3$ and related compounds the author concluded that the notation of the YAlO$_3$ crystal axes was permuted in Ref.~\onlinecite{Aggarwal-2005}. This suggestion is in fact in perfect agreement with our finding as demonstrated in the inset of Fig. \ref{YAlO_kappa}. It compares the data from Ref.~\onlinecite{Aggarwal-2005}, %\textcolor{red}{
however,%} 
with the orientation labeled according to Ref.~\onlinecite{Hofmeister-2010} to our thermal conductivity. Both data sets have similar values and anisotropy. In addition we plot the high-temperature results from Ref.~\onlinecite{Zhang-2023}, which have somewhat larger absolute values but a similar anisotropy and $T$ dependence.

The remaining differences between results from different groups may be due to measurement uncertainties, especially for sample dimensions. As mentioned in the experimental section, our geometry factor for YAlO$_3$ and in consequence absolute $\kappa$ values are only known within $\pm 11 \%$. Uncertainties from other sources may be neglected in our case. On the other hand, the thermal conductivity data in Ref.~\onlinecite{Aggarwal-2005,Zhang-2023} were obtained from diffusivity and specific heat measurements, the accuracy of which is estimated in Ref.~\onlinecite{Aggarwal-2005} to be $7 \%$ for each quantity. The respective data sets can be scaled to ours by a factor of roughly 0.77, which is within the combined uncertainty of absolute values. The deviations of the data in Ref.~\onlinecite{Zhang-2023} are much larger, however no details on the measurement accuracy are specified. %In combination with the relevance of sample purity for transport measurements, these effects may account for the observed deviations between the results of different groups.

In this context one may ask, whether the observed anisotropy $\kappa_c / \kappa_b \approx 1.35$ is an established fact or whether it might be attributed to measurement uncertainties as well. The range of $\pm 11 \%$ for our absolute $\kappa$ values for an individual sample corresponds to an uncertainty of $22 \%$ for the ratio between two different measurements. However, this value represents the worst case and is unlikely to be realized. Moreover, it is still too small to fully account for the difference between $\kappa_b$ and $\kappa_c$. This is demonstrated in Fig.~\ref{anisotropy}, which compares the ratio $\kappa_c/\kappa_b$ for different data sets including estimates of the maximum uncertainty range for selected data points. The isotropic value $\kappa_c / \kappa_b = 1$ is marked by a dashed line and lies below the uncertainty range for YAlO$_3$ except for a small window around 20 K. Moreover, the anisotropy ratio $\kappa_c / \kappa_b$ strongly increases below 10~K to about 1.8 at 2 K. Such a temperature dependence cannot be attributed to measurement uncertainties of the contact geometry. In combination with the very similar anisotropy observed in previous studies, this allows us to conclude that the thermal conductivity of YAlO$_3$ indeed exhibits a moderate anisotropy $\kappa_c / \kappa_b$ with a weak temperature dependence below room temperature.

%%%%%%%%%%%%%%%%%%%%%%

\subsection{Heat transport in YbAlO$_3$}

\begin{figure}[t!] %{wrapfigure}{r}{0.55\textwidth}
\centering 
 \includegraphics[width=0.6\columnwidth]{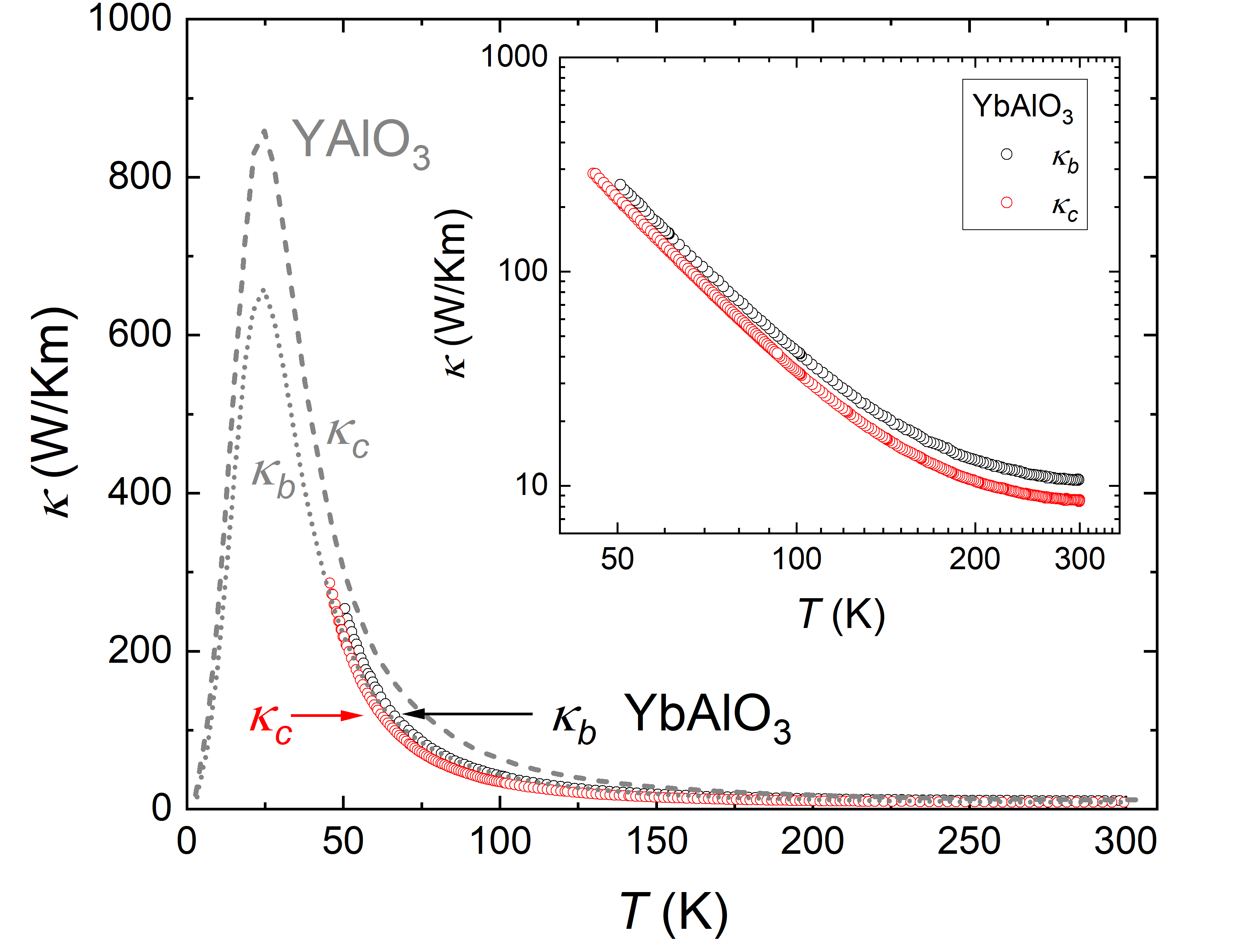}
\caption{Temperature dependence of the thermal conductivity \(\kappa\) of YbAlO$_3$ 
along the \(b\)-axis (\(\kappa_b\)) and the \(c\)-axis (\(\kappa_c\)) plotted in black and red, respectively. Data for YAlO$_3$ are shown as dashed lines for comparison. The inset presents the YbAlO$_3$ data on a double-logarithmic scale. %including $0.8 \kappa_b$ (blue line) to demonstrate the scaling of the thermal conductivity curves.
}
\label{YbAlO_kappa}
\end{figure}

The thermal conductivity of YbAlO$_3$ is shown in Fig.~\ref{YbAlO_kappa}. For comparison, the data of YAlO$_3$ are added as lines. The inset of Fig.~\ref{YbAlO_kappa} plots the same data on a double logarithmic scale. As mentioned in the experimental section, measurements on YbAlO$_3$ were not successful below about 50~K due to the large thermal conductivity and the small crystal size. The behavior of $\kappa_b$ and $\kappa_c$ of YbAlO$_3$ is very similar to the one of YAlO$_3$, both for the $T$-dependence and the absolute values. The latter observation is somewhat unexpected in view of the considerable difference of the minimum thermal conductivities $\kappa_{\mathrm{min}}$ predicted from density functional theory (DFT) calculations of elastic constants and the model proposed by Clarke, namely 1.61 W/Km for YAlO$_3$ compared to 1.15 W/Km for YbAlO$_3$~\cite{zhan_theoretical_2012, Xiang-2015, Clarke-2003}. However, the regime of $\kappa_{\mathrm{min}}$ is only reached at temperatures well above the Debye temperature $\Theta_\mathrm{D}$, i.e. far above our measurement range. 

\begin{figure}[t!] %{wrapfigure}{r}{0.55\textwidth}
\centering 
 \includegraphics[width=0.6\columnwidth]{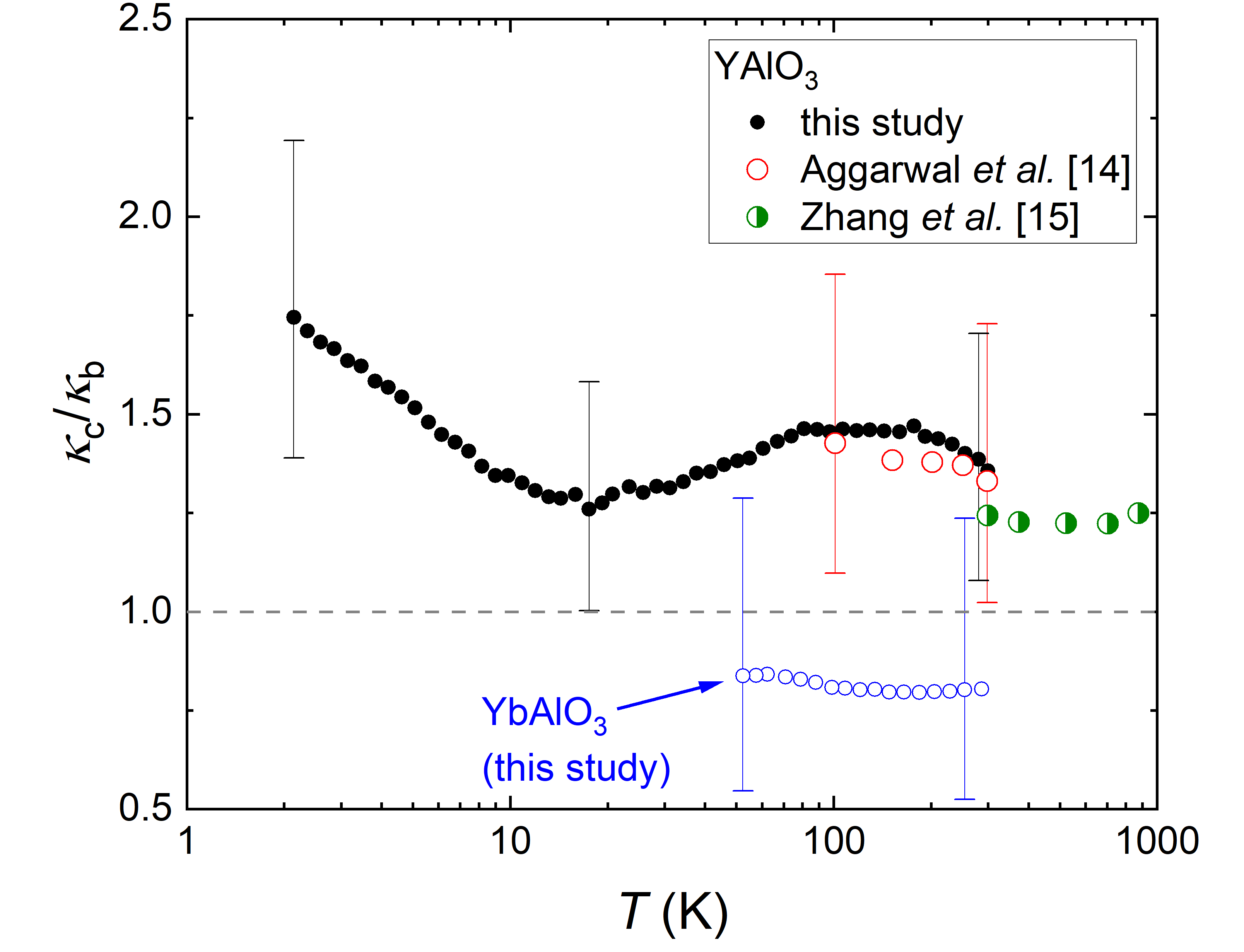}
\caption{Temperature dependence of the thermal conductivity ratio $\kappa_c/\kappa_b$ of YAlO$_3$ and YbAlO$_3$ including literature data from Aggarwal \textit{et al.} \cite{Aggarwal-2005} and Zhang \textit{et al.} \cite{Zhang-2023} on YAlO$_3$. The error bars demonstrate maximum uncertainties from the upper and lower bounds for the geometry factors and measurement techniques. No respective estimate is provided by Zhang \textit{et al.}}
\label{anisotropy}
\end{figure}

The very similar magnitude and temperature dependence of $\kappa$ for YbAlO$_3$ and YAlO$_3$ in the $T$ range above 50~K points to a rather uniform behavior also at lower $T$. Therefore, we suspect, that YbAlO$_3$ is a very good thermal conductor down to temperatures of liquid He with a maximum in $\kappa$ around 20-30 K. This assumption is also supported by the fact that we could not recover a sizable thermal gradient at lower $T$ in our setup, meaning that the thermal conductivity stays indeed large down to 2~K. However, absolute values at low $T$ cannot be evaluated from high-$T$ data, because different scattering processes dominate in the respective regimes. This is discussed in more detail below.

Another interesting point is the anisotropy of $\kappa$. For YbAlO$_3$ $\kappa_c < \kappa_b$, which is opposite to the anisotropy for YAlO$_3$. %The scaling $\kappa_c \approx 0.8 \kappa_b$ is illustrated in the inset of Fig.~\ref{YbAlO_kappa}. 
The temperature dependent ratio $\kappa_c/\kappa_b$ of YbAlO$_3$ is shown in Fig.~\ref{anisotropy} including estimates for the maximum uncertainty range of absolute values. From this plot it is clear, that a discussion of the anisotropy of $\kappa$ of YbAlO$_3$ is not reasonable based on our data because of the uncertainty of the geometry factors. While YbAlO$_3$ has most likely a smaller ratio $\kappa_c/\kappa_b$ than YAlO$_3$ above 50~K with an only weak $T$ dependence, we cannot draw any conclusion about the absolute values, in particular whether $\kappa_c/\kappa_b$ is larger or smaller than 1.

%%%%%%%%%%%%%%%%%%%%%%

\subsection{Discussion}
As mentioned above, the temperature dependencies of $\kappa (T)$ for YAlO$_3$ and YbAlO$_3$ are typical for electrically insulating single crystals. In the following we discuss the data in more detail focusing on the origin of the anisotropy of $\kappa$ and its temperature dependence in YAlO$_3$. %In the following we discuss the data in more detail focussing on two questions: (1) What is the origin of the anisotropy of $\kappa$ and its temperature dependence, especially in YAlO$_3$. And (2) Why is the effect of Y-Yb exchange so small despite the large difference in density. 

A simple scenario to explain a thermal transport anisotropy is a direction-dependence of the sound velocity $v$. Starting with the kinetic relation $\kappa = 1/3 \, c_V v \lambda = 1/3 \, c_V v^2 \tau$, with $c_V$ being the specific heat at constant volume, one may assume either an isotropic mean free path $\lambda$ yielding an anisotropy $\kappa _c / \kappa _b = v_c / v_b$ or an isotropic scattering time $\tau$ corresponding to $\kappa _c / \kappa _b = v_c^2 / v_b^2$. Since there is no experimental report on the direction-dependent velocities of sound of neither YAlO$_3$ nor YbAlO$_3$, we use two routes for an estimation: (1) from the elastic constants and (2) from phonon dispersion curves. For the first estimation we take the elastic constants calculated by an \textit{ab initio} DFT method by Zhan \textit{et al.}(YAlO$_3$) \cite{zhan_theoretical_2012} and Xiang \textit{et al.} (YbAlO$_3$) \cite{Xiang-2015} to compute the 
sound velocities along different principal axes. For each crystallographic direction of \(i j k\) (e.g. the \(i\)-axis), one may define one longitudinal velocity $v_l$ ($v_{ii}$) and two transverse velocities of $v_{t1}$ ($v_{i,j}$) and $v_{t2}$ ($v_{i,k}$). For an orthorhombic crystal structure, the relation to the elastic constants reads: \cite{Martelli-2021} 
\begin{equation}\label{eq:velocity_matrix}
     \hat{v} = \begin{pmatrix}
  v_{i,i} & v_{i,j} &  v_{i,k} \\
  v_{j,i} & v_{j,j} &  v_{j,k}  \\
  v_{k,i} & v_{k,j} &  v_{k,k} 
 \end{pmatrix} 
 = \frac{1}{\sqrt{\rho}} \begin{pmatrix}
  \sqrt{c_{11}} & \sqrt{c_{66}} &  \sqrt{c_{55}} \\
  \sqrt{c_{66}} & \sqrt{c_{22}} &  \sqrt{c_{44}}  \\
  \sqrt{c_{55}} & \sqrt{c_{44}} &  \sqrt{c_{33}} 
 \end{pmatrix} 
\end{equation}
where \(\rho\) is the density, \(c_{11}\), \(c_{22}\), and \(c_{33}\) represent stiffness against uni-axial strains, and \(c_{44}\), \(c_{55}\), and \(c_{66}\) correspond to shear deformations. Once the longitudinal and transverse velocities are known, an estimate for the mean sound velocity for a specific direction can be found via $\overline{v}_i^{-3} = ({v_{i,i}}^{-3} + {v_{i,j}}^{-3} + {v_{i,k}}^{-3} ) /3 $, which is strictly valid only for isotropic materials. The results are given in Table \ref{tab:velocities}, together with the overall average sound velocities $\overline{v}$ calculated as the geometric mean of the three directions and the thermal conductivity anisotropies determined as $\kappa _c/ \kappa _b = \overline{v}_c^2 / \overline{v}_b^2$.

\begin{table}
\caption{\label{tab:velocities}%
Mean sound velocities for YAlO$_3$ and YbAlO$_3$ along the three crystallographic axes calculated from the elastic tensor $\epsilon$ or phonon dispersion curves (PhDis) and resultant overall average sound velocity $\overline{v}$ and thermal conductivity anisotropy $\kappa _c/ \kappa _b = \overline{v}_c^2 / \overline{v}_b^2$.}
\begin{ruledtabular}
\begin{tabular}{ccccccc}
    Material & source & $\overline{v}_a$ & $\overline{v}_b$ & $\overline{v}_c$ & $\overline{v}$ &  $\kappa _c/ \kappa _b$ \\
    & & (m$/$s) & (m$/$s) & (m$/$s) & (m$/$s) & \\
    \colrule
    YAlO$_3$ & $\epsilon$ \cite{zhan_theoretical_2012} & 6020 & 5410 & 5460 & 5640 & 1.02 \\
    YAlO$_3$ & PhDis \cite{Suda-2003} & 6110 & 4920 & 5510 & 5530 & 1.25 \\
    YAlO$_3$ & PhDis \cite{TOGO-1} & 5120 & 5220 & 5750 & 5370 & 1.21 \\
    YbAlO$_3$ & $\epsilon$ \cite{Xiang-2015} & 3890 & 3930 & 4410 & 4080 & 1.26 \\
\end{tabular}
\end{ruledtabular}
\end{table}

The second method can be applied only for YAlO$_3$, since no phonon dispersion curves have been published for the Yb compound. Theoretical phonon dispersion curves for YAlO$_3$ at room temperature have been obtained from a rigid ion model and Raman results~\cite{Suda-2003} as well as from ab initio calculations using the VASP and phonopy codes~\cite{TOGO-1}. The sound velocities calculated from the slopes of the acoustic branches and resultant values for $\overline{v}$ and $\kappa _c/ \kappa _b$ are also presented in Table~\ref{tab:velocities}.

We start with the results for YAlO$_3$: Two findings are interesting with respect to our thermal conductivity measurements. First, the overall average sound velocities $\overline{v}$ are close to each other for all methods of calculation and only slightly larger than the polycrystalline average sound velocity of 5290~m$/$s calculated by Zhan \textit{et al.} based on the Voigt, Reuss, and Hill approximations~\cite{zhan_theoretical_2012, Voigt-1928, Reuss-1929, Hill-1952}. Second, the anisotropy of the sound velocities depends on the method of estimation. In particular, the calculation from the elastic constants reveals an almost isotropic behavior along $b$ and $c$ and slightly larger value along the $a$-axis. This is even more remarkable in view of the considerable anisotropy of the elastic constants, e.g. $c_{cc}/ c_{bb} = c_{33}/ c_{22} = 0.78$. Apparently, they cancel out when averaging over longitudinal and transverse components. As result, the ratio of the mean sound velocities calculated from the elastic constants cannot explain the observed thermal transport anisotropy of YAlO$_3$, neither for an isotropic mean free path ($\overline{v}_c / \overline{v}_b = 1.01$) nor an isotropic scattering time ($\overline{v}_c^2 / \overline{v}_b^2 = 1.02$). We attribute this to the approximations used in the model. By contrast the sound velocities obtained from the phonon dispersion curves are in line with a larger thermal conductivity along $c$ than along $b$, and the anisotropies $\kappa _c/\kappa _b = \overline{v}_c^2 / \overline{v}_b^2$ of about 1.2 calculated assuming an isotropic $\tau$ are comparable to the experimental one, although the temperature dependence cannot be explained by this simple estimate. %and discuss this point in more detail below. At the end, we also demonstrate that the anisotropy of $\kappa$ is nonetheless due to the one of $v$.  

Next we look at YbAlO$_3$. All calculated sound velocities are considerably smaller than for YAlO$_3$. This is in contradiction to our experimental finding of similar $\kappa$ values for both compounds given the simple proportionality $\kappa \propto v$ or $\kappa \propto v^2$. Moreover, the anisotropy $\kappa _c/ \kappa _b$ for YbAlO$_3$ predicted by the velocities is opposite to our results. 
% Altogether, our thermal conductivity results with their anisotropy, temperature dependence, and the change in $\kappa$ upon Y-Yb substitution cannot be explained by a simple variation in the sound velocities calculated from the elastic constants. 
While part of the discrepancies may be caused by the considerable uncertainty of our absolute $\kappa$ values, the applied simplifications of reducing the full phonon dispersion relation to three sound velocities are probably to coarse and additional parameters, as a temperature and orientation dependence of the mean free path or anharmonic effects need to be taken into consideration. For instance, the above mentioned DFT calculations also predict a significant difference of the Gr\"uneisen parameter between the Y and Yb compound~\cite{zhan_theoretical_2012, Xiang-2015}. This parameter is used in  describing anharmonic scattering processes and can have a considerable effect on $\kappa$~\cite{Slack-1964}. However, the calculated values represent a high-$T$ limit and are thus of limited validity in our investigated $T$ range. Likewise, the elastic constants used in the estimation of the sound velocities are susceptible to the approximations and simplifications immanent to the applied model. Most important, the DFT calculations are performed for $T = 0$ and cannot describe any $T$-dependence.

\begin{figure}[t!] %{wrapfigure}{r}{0.55\textwidth}
\centering 
 \includegraphics[width=0.6\columnwidth]{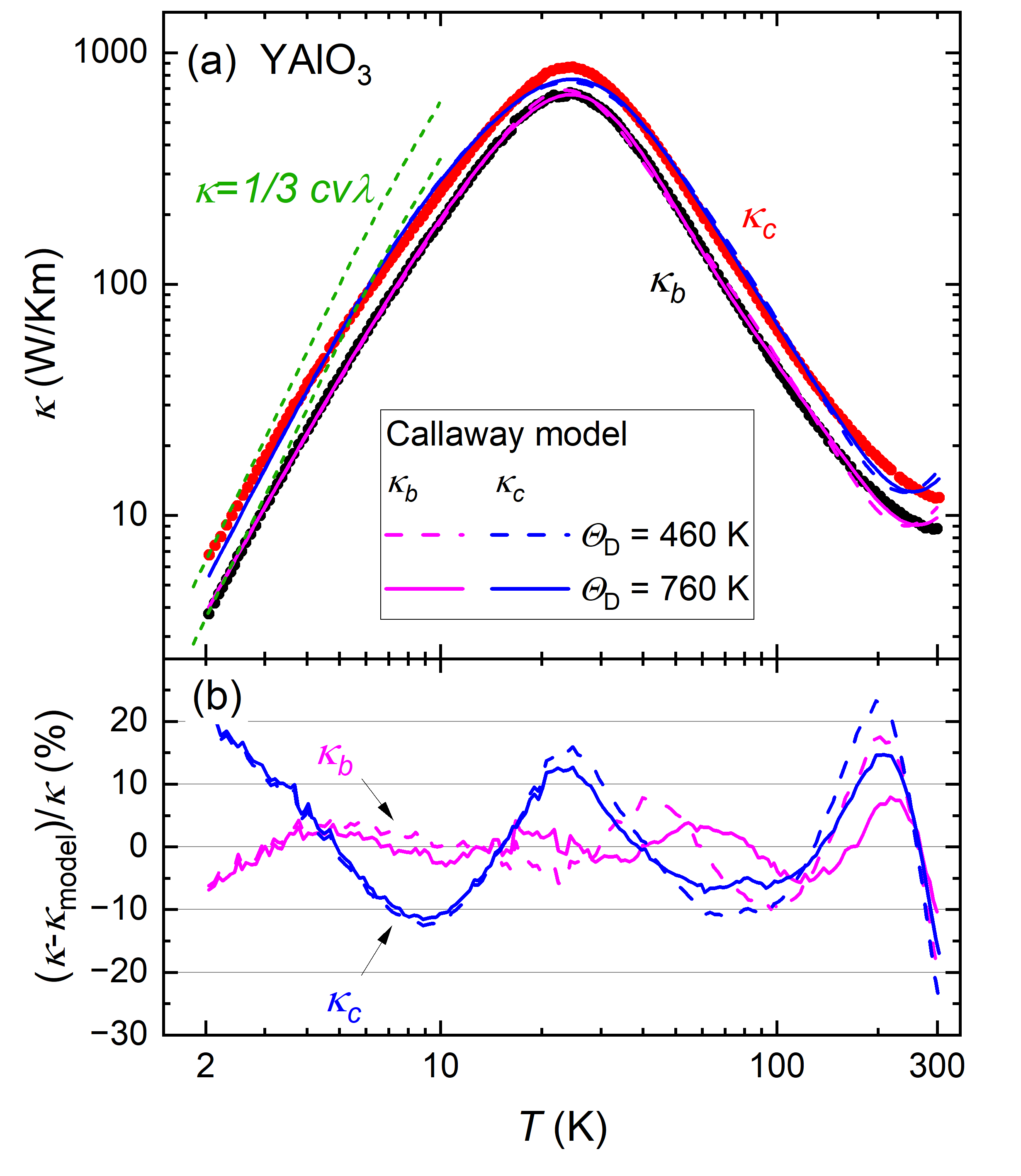}
\caption{(a) The upper panel shows the results of modeling the thermal conductivities $\kappa_b$ and $\kappa_c$ of YAlO$_3$ by the Callaway model. The calculated curves for $\kappa_b$ can be barely  distinguished from the data and from each other, except for the range $T>200$~K. For $\kappa_c$, systematic deviations between model and data are seen at all $T$. Details of the calculation procedure are given in the main text. The corresponding parameters are summarized in Table~\ref{tab:YALO_fit}. The dotted lines at low $T$ demonstrate the behaviors expected from the kinetic relation for constant mean-free paths $\lambda$. (b) The lower panel shows the deviations between model and data, emphasizing the much better agreement between for $\kappa_b$ than for $\kappa_c$.}
\label{callaway}
\end{figure}

In order to understand the temperature dependence of the thermal conductivity anisotropy, different scattering processes and their $T$ dependence have to be taken into account. At very low temperatures, phonons are predominantly scattered by sample or grain boundaries. In first approximation, this regime is characterized by a constant mean free path $\lambda$ and a cubic $T$ dependence of $\kappa$. For YAlO$_3$ $\lambda$ can then be determined from the kinetic relation using specific heat data and an average value for the sound velocity of 5500 m$/$s. Fig.~\ref{callaway} shows the YAlO$_3$ data again in a double logarithmic presentation together with dotted lines illustrating the respective behavior. The thermal conductivity range, which may be approximated by this simple relation is limited to below at most 4~K. Our estimates for the phonon mean free paths at low $T$ for $j_Q \parallel b$ and $c$ are $17 \, \mathrm{\mu m}$ and $29 \, \mathrm{\mu m}$, respectively. These values are about one order of magnitude below the minimum sample dimension indicating that thermal transport at low $T$ is limited by defects and not the sample size. Moreover, the rather poor agreement between a simple cubic behavior and our data demonstrates, that other scattering centers besides boundaries are relevant. Nevertheless, the rather large values for the phonon mean free path at low $T$ confirm the high purity of the investigated single crystals.

At higher temperatures, additional scattering mechanisms become relevant. A phenomenological model frequently used to analyze the relevance of different scattering processes in the thermal conductivity has been proposed by Callaway~\cite{Callaway-1959}. This model is based on the relaxation time and Debye approximations. The original version does not take into account dispersion and anisotropies of the sound velocity or phonon spectrum. It has been subsequently expanded to deal with more complex situations, see e.g., Ref.~\onlinecite{Holland-1963, Palmer-1997, Allen-2013}. In order to limit the number of free parameters, we stick to the original variant of this model, assuming an isotropic velocity of sound. The total lattice thermal conductivity is determined from a combined relaxation time $\tau_{\mathrm{c}}$ defined as $\tau_{\mathrm{c}}^{-1} = \tau_{\mathrm{q}}^{-1} + \tau_{\text{N}}^{-1}$ with $\tau_{\mathrm{q}}$ and $\tau_{\mathrm{N}}$ as the phonon scattering times via resistive and nonresistive (normal) 3-phonon-processes, respectively. Although nonresistive, normal processes do not limit thermal conduction directly, they are relevant as they affect the phonon distribution function. 
The temperature and frequency dependence of $\tau_{\mathrm{N}}$ is calculated from $\tau^{-1}_{\mathrm{N}}(\omega, T) = N T^\alpha \omega ^\beta$ with $\alpha = 2$, $\beta = 3$, and $N$ as a free parameter. $\tau_{\mathrm{q}}$ in turn is determined from all individual resistive scattering rates. In our case we consider scattering at the sample boundaries ($\tau_{\mathrm{b}}$), at point defects ($\tau_{\mathrm{def}}$), at dislocations ($\tau_{\mathrm{dis}}$), and due to phonon-phonon U-scattering processes ($\tau_{\mathrm{U}}$). Thus \(\tau_{\mathrm{q}}^{-1} = \tau_{\mathrm{b}}^{-1} + \tau_{\mathrm{def}}^{-1} + \tau_{\mathrm{dis}}^{-1} + \tau_{\mathrm{U}}^{-1}\), which reads in more detail:
\begin{equation}
\label{eq:YALO_fit_phonon}
    \tau^{-1}_{\mathrm{q}} (\omega,T) = \frac{\overline{v}}{l} + A \omega^4+ D \omega + B T^\alpha \omega^\beta \exp(-\frac{\Theta _\mathrm{D}}{b T}) 
\end{equation} 
with the terms representing the individual scattering times in the respective order. $\overline{v}$ and $l$ are the average velocity of sound and the mean sample cross-section dimension. The exponents $\alpha$ and $\beta$ are the same as for normal processes. \(A\), \(D\), \(B\), and \(b\) are free parameters of the model. In our calculation of $\kappa$ we also include a contribution with cubic $T$ dependence $cT^3$ for possible radiation losses at high $T$.

Some values of equation \ref{eq:YALO_fit_phonon} can be determined prior to modeling. As mean sample cross-section dimension we use $l = \sqrt{4 A_s/\pi}$ with the cross-section $A_s$. For the average sound velocity $\overline{v}$ we take a value of $5500$~m$/$s corresponding to an average from the elastic constants and the phonon dispersion estimates. The situation is more difficult for the Debye temperature $\Theta_\mathrm{D}$ as literature values vary significantly. The shift of phonon lines in a photoluminescence study yielded $\Theta_\mathrm{D} = 465$~K~\cite{Zhydachevskii-2006}. Fits of the temperature-dependence of the lattice parameters $a$ and $c$ resulted in values of 440~K and 466~K with rather large uncertainties of about $\pm 90$~K, while the respective value for the $b$ lattice parameter of about 900~K is most probably overestimated~\cite{Senyshyn-2013}. The corresponding evolution of the unit-cell volume yielded $\Theta_\mathrm{D} =545$~K. From specific heat measurements a Debye temperature of 760~K was determined~\cite{Aggarwal-2005}. Balancing the large spread of literature values for $\Theta_\mathrm{D}$ and our aim to reduce the number of parameters in equation \ref{eq:YALO_fit_phonon} we decided to model our thermal conductivities for two fixed values, namely 460~K and 760~K. This procedure also allows some estimation of the stability of the other parameters.

Having reduced the number of free parameters from 9 to 6 we modeled our thermal conductivity data for YAlO$_3$ by sampling the parameter space incrementally. As criterion for the agreement between model and data we used the root mean square deviation RMSD in a double logarithmic presentation. In order to consider the whole measurement range with equal relevance, we averaged our original data set with a step width of 3\% in temperature. This procedure yielded the same number of data points for both orientations.

Fig.~\ref{callaway}a shows the best descriptions of our data sets for $\Theta_\mathrm{D} = 460$~K and $\Theta_\mathrm{D}= 760$~K. The corresponding model parameters and RMSD values are summarized in Table~\ref{tab:YALO_fit}. The uncertainties indicate the width of 2 RMSD determined by separately varying each parameter. I.e., these values do not take into account mutual dependencies and are merely meant to give an impression of the parameter precision. In some cases, only an upper or lower limit is specified. A missing lower limit for $A$ and  $N$ means that the respective scattering process can be ignored without reducing the fit quality significantly. The absence of an upper limit for $N$ implies that even infinitely rapid normal processes are in line with the data.

\begin{figure}[t!] %{wrapfigure}{r}{0.55\textwidth}
\centering 
\includegraphics[width=0.6\columnwidth]{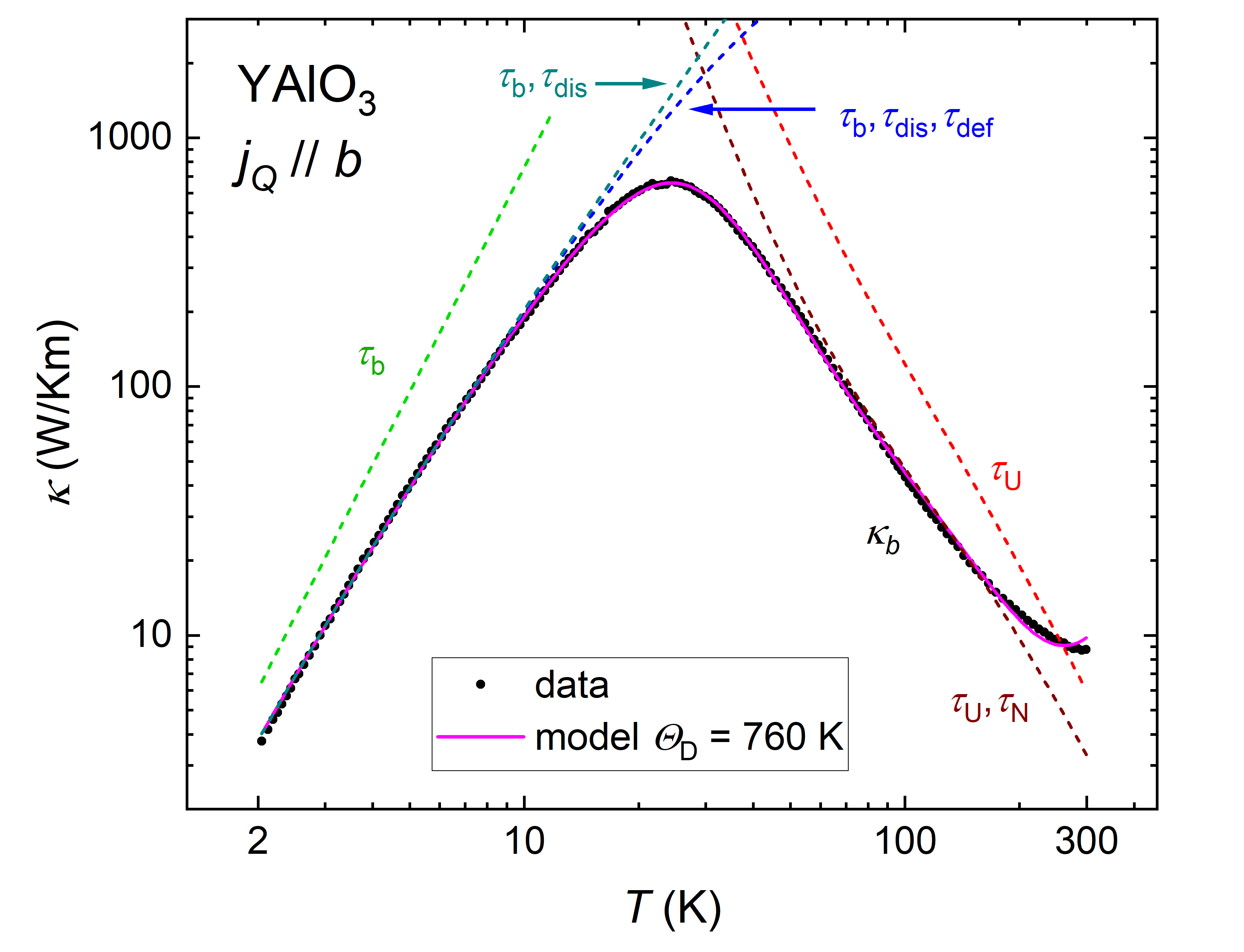}
\caption{Influence of the different scattering contributions on the thermal conductivity of YAlO$_3$ for $j_Q \parallel b$ using $\Theta _\mathrm{D} = 760$~K. The dashed lines correspond to situations when only the specified relaxation times are taken into account.}
\label{contributions}
\end{figure}

\begin{table*}
\caption{\label{tab:YALO_fit}
Best parameters found to describe the \(T\) dependence of $\kappa$ for YAlO$_3$ in a double-logarithmic presentation by the Callaway model. The root mean square deviation RMSD is a measure for the deviation of model and data.}
\begin{ruledtabular}
\begin{tabular}{ccccccccc}
Sample & $\Theta_\mathrm{D}$ & $N$ & $A$ & $D$ & $B$ & $b$ & $c$ & RMSD \\ 
& (K) & ($10^{-22}$ s/K$^{3}$) & ($10^{-46}$ s$^3$) & $10^{-6}$ & ($10^{-23}$ s/K$^{3}$) & & $10^{-7}$ W/mK$^{4}$ & \\ 
& & normal & point & dislocations & umklapp & umklapp & radiation &  \\ 
& & processes & defects & & scattering & scattering & losses &  \\ 
\colrule
    YAlO$_3$ \#b & 460 & 25 $$(> 3)$$ & 0.7 $(< 3.7)$ & $5.2 \pm 1.3$  & $6.7 \pm 1.3$ & $12.3 \pm 3.8$ & 3.1 & 0.026 \\
    YAlO$_3$ \#b & 760 & 8 $(4-40)$ & $3.9 \pm 1.5$ & $5.1 \pm 0.7$ & $8.2 \pm 1.0$ & $12.1 \pm 1.8$ & 2.4 & 0.013 \\
    YAlO$_3$ \#c & 460 & $< 0.4$ & 80 $(20-180)$ & $1.5 \pm 1.4$ & $9.0 \pm 3.4$  & 13.5  $(7-37)$ & $4.5$ & 0.046 \\
    YAlO$_3$ \#c & 760 & $< 0.4$ & 80 $(30-160)$ & $1.5 \pm 1.2$ & $11.0 \pm 3.2$ & 17.6 $(12-30)$ & 3.7 & 0.037 \\
\end{tabular}
\end{ruledtabular}
\end{table*}

For $\kappa_b$, the agreement between calculated curves and data is very good given the simplicity of the model. In fact, the model curves can be barely distinguished from the data and from each other on the scale of Fig.~\ref{callaway}a, except for the region above about 200 K. This is also seen in Fig.~\ref{callaway}b showing the relative difference between data and model. In a large $T$ range it is well below 10~\%, exceeding this value only above 150~K. These deviations at high $T$ cannot be explained satisfactorily by radiation losses with cubic temperature dependence, an effect, which was included in our model and is responsible for the upturn in the calculated curves at highest $T$. Instead, we suspect that the Debye approximation is no longer a good description of the phonon spectrum in this $T$ range, for instance due to an increasing contribution from anharmonic effects or optical modes. The latter is in line with the observation of a low-lying Einstein mode in thermal expansion measurements~\cite{Senyshyn-2013}. The characteristic Einstein temperature was found to be about $230$~K $\pm 130$~K, i.e., it might be relevant for the thermal conductivity above 200~K.

In contrast to $\kappa_b$, $\kappa_c$ could not be modeled satisfactorily. The calculated curves show systematic deviations from the data, independent of the choice of $\Theta_\mathrm{D}$. This is also visible from the relative deviation between data and model for $\kappa_c$ in Fig.~\ref{callaway}b, which is much larger than for $\kappa_b$ and oscillates in the whole investigated $T$ range meaning that there is a systematic deviation of the data from the model. Moreover, the parameter set for the best description is almost the same for $\Theta_\mathrm{D} = 460$~K and 760~K, cf.~Table~\ref{tab:YALO_fit}. Either additional scattering mechanisms play a role, or the Debye approximation and an isotropic sound velocity are not compatible with the observed anisotropy of the thermal conductivity of YAlO$_3$.

The influence of the different scattering contributions on the thermal conductivity of YAlO$_3$ is demonstrated examplarily for $\kappa_b$ and $\Theta _\mathrm{D} = 760$~K in Fig.~\ref{contributions}. The low-$T$ part of $\kappa$ is defined by a combination of scattering from sample boundaries and dislocations. At high $T$ the thermal conductivity is predominantly limited by normal and umklapp scattering processes. Point defects are of minor importance in the whole investigated temperature range. In particular, the mean free path at lowest $T$ estimated above to be $\approx 20 \, \mu$m is mostly limited by dislocations. %\textcolor{red}{
It may be somewhat unexpected that the corresponding fit parameter for defect scattering is anisotropic. However, we may not exclude, that our YAlO$_3$ crystals stem from different batches with a different impurity concentration. At this point we would also like to recall, that there is a considerable mutual dependence between the fit parameters for scattering processes dominating at low $T$, namely from boundaries, point defects, and dislocations, and for those relevant at high $T$, i.e. umklapp and normal scattering processes.%}

It is not reasonable to model the thermal conductivity of YbAlO$_3$ in the same way as for YAlO$_3$. Our calculations for YAlO$_3$ reveal a poor reproducibility of the data by the model above 200~K, most probably due to limitations of the Debye approximation. In combination with the reduced measurement region for YbAlO$_3$ the $T$ range for modeling is simply to small to get reliable results for this material. This precludes also an extrapolation of our data to very low temperatures to get a good estimate of the phonon contribution to the thermal conductivity in the milli-Kelvin region. A different approach has to be taken in order to disentangle the magnetic and lattice thermal conductivities in YbAlO$_3$ in the quasi-one-dimensional quantum magnet regime below 1~K.

Finally we return to our motivation of gaining information on the low-temperature thermal conductivity of YAlO$_3$ and the effect of rare-earth ion substitution. YAlO$_3$ exhibits a very good thermal conductivity down to liquid helium temperatures. We found a moderate anisotropy with $\kappa_b > \kappa_c$. This allows in principle optimizing thermal properties for applications by choosing an appropriate crystal orientation. However, the influence of anisotropy is probably of minor importance compared to other aspects as realization of a good thermal coupling and crystal quality. 

Replacement of Y by much heavier Yb has only a small effect on the thermal conductivity above 50~K, i.e. the regime, where umklapp scattering is dominating. An extrapolation of the measured thermal conductivity of our YbAlO$_3$ crystals towards lower $T$ is not possible, as it depends sensitively on crystal size and purity. However, for Y- and Yb crystals of comparable quality the similarity above 50 K both in absolute values and $T$ dependence of $\kappa$ suggests that it extends also to lower $T$, as long as no magnetic contributions from Yb moments in scattering or transport become important. These effects are expected to play a role only at very low $T$ due to the localized character of the Yb magnetic moments with an ordering temperature of below 1~K~\cite{Wu-2019}.

\section{Summary} 

We have measured large thermal conductivities with weak anisotropy in single crystals of YAlO$_3$ and YbAlO$_3$. The exchange of Y by Yb has a minor effect on $\kappa$ down to 50~K despite the considerable change in density and average atomic mass. Our results on YAlO$_3$ confirm the suitability of the material for applications requiring a low thermal resistance, e.g., in lasers, sensors, and as substrate materials at temperatures down to liquid He.

\begin{acknowledgments}
We wish to acknowledge funding from the Deutsche Forschungsgemeinschaft (DFG, German Research Foundation) through the Collaborative Research Centers  SFB 1143 and TRR 80 and through the W\"urzburg-Dresden Cluster of Excellence EXC
2147 on Complexity and Topology in Quantum Materials - ct.qmat. We acknowledge financial support from the Max Planck Society through the Physics of Quantum Materials department and the research group
''Physics of Unconventional Metals and Superconductors
(PUMAS)''.
\end{acknowledgments}

\section*{Data Availability Statement}

 The data that support the findings of this article are openly
 available \cite{data}.

\bibliography{YAO-refs}

@article{Suda-2003,
  title = {The first-order {Raman} spectra and lattice dynamics for {YAlO$_3$ crystal}},
  author = {Suda, Jun and Kamishima, Osamu and Hamaoka, Kohhei and Matsubara, Ichirou and Hattori, Takeshi and Sato, Tsutomu},
  journal = {J.~Phys.~Soc.~Jpn.},
  volume = {72},
  issue = {6},
  pages = {1418},
  year = {2003},
  doi = {10.1143/JPSJ.72.1418},
  url = {https://link.aps.org/doi/10.1143/JPSJ.72.1418}
}

@article{Robles-2003,
title = {On the possibility of growing unidirectionally twinned {YBa$_2$Cu$_3$O$_{7 -  \delta}$ thin films on YAlO$_3$}},
journal = {Physica C},
volume = {400},
number = {1},
pages = {36-42},
year = {2003},
doi = {10.1016/S0921-4534(03)01327-3},
url = {https://www.sciencedirect.com/science/article/pii/S0921453403013273},
author = {J. J. Robles and A. Bartasyte and H. P. Ng and A. Abrutis and F. Weiss},
}

@article{Slack-1964,
  title = {Thermal Conductivity and Phonon Scattering by Magnetic Impurities in {CdTe}},
  author = {Slack, Glen A. and Galginaitis, S.},
  journal = {Phys. Rev.},
  volume = {133},
  issue = {1A},
  pages = {A253--A268},
  numpages = {0},
  year = {1964},
  month = {Jan},
  publisher = {American Physical Society},
  doi = {10.1103/PhysRev.133.A253},
  url = {https://link.aps.org/doi/10.1103/PhysRev.133.A253}
}

@article{Sun-2025,
author = {Fei Sun  and Simli Mishra  and Ulrike Stockert  and Ramzy Daou  and Naoki Kikugawa  and Robin S. Perry  and Elena Hassinger  and Sean A. Hartnoll  and Andrew P. Mackenzie  and Veronika Sunko },
title = {The {Lorenz} ratio as a guide to scattering contributions to transport in strongly correlated metals},
journal = {Proceedings of the National Academy of Sciences},
volume = {121},
number = {35},
pages = {e2318159121},
year = {2024},
doi = {10.1073/pnas.2318159121},
URL = {https://www.pnas.org/doi/abs/10.1073/pnas.2318159121},
eprint = {https://www.pnas.org/doi/pdf/10.1073/pnas.2318159121},
}

@article{Behnia-2018,
  title = {Thermal Transport and Phonon Hydrodynamics in Strontium Titanate},
  author = {Martelli, Valentina and Jim\'enez, Julio Larrea and Continentino, Mucio and Baggio-Saitovitch, Elisa and Behnia, Kamran},
  journal = {Phys. Rev. Lett.},
  volume = {120},
  issue = {12},
  pages = {125901},
  numpages = {6},
  year = {2018},
  month = {Mar},
  publisher = {American Physical Society},
  doi = {10.1103/PhysRevLett.120.125901},
  url = {https://link.aps.org/doi/10.1103/PhysRevLett.120.125901}
}

@article{Uchida-2017,
  title = {Samarium monoxide epitaxial thin film as a possible heavy-fermion compound},
  author = {Uchida, Yutaka and Kaminaga, Kenichi and Fukumura, Tomoteru and Hasegawa, Tetsuya},
  journal = {Phys. Rev. B},
  volume = {95},
  issue = {12},
  pages = {125111},
  numpages = {4},
  year = {2017},
  month = {Mar},
  publisher = {American Physical Society},
  doi = {10.1103/PhysRevB.95.125111},
  url = {https://link.aps.org/doi/10.1103/PhysRevB.95.125111}
}

@article{Wu-2019,
	author = {Wu, L. S. and Nikitin, S. E. and Wang, Z. and Zhu, W. and Batista, C. D. and Tsvelik, A. M. and Samarakoon, A. M. and Tennant, D. A. and Brando, M. and Vasylechko, L. and Frontzek, M. and Savici, A. T. and Sala, G. and Ehlers, G. and Christianson, A. D. and Lumsden, M. D. and Podlesnyak, A.},
	title = {{Tomonaga-Luttinger} liquid behavior and spinon confinement in  {YbAlO$_3$}},
	volume = {10},
	issn = {2041-1723},
	url = {https://doi.org/10.1038/s41467-019-08485-7},
	doi = {10.1038/s41467-019-08485-7},
	number = {1},
	journal = {Nat. Commun.},
	month = feb,
	year = {2019},
	pages = {698},
}

@article{Dewo-2021,
author = {Dewo, Wioletta and Gorbenko, Vitaliy and Syrotych, Yurii and Zorenko, Yuriy and Runka, Tomasz},
title = {Mn-Doped {XAlO$_3$ (X = Y, Tb)} Single-Crystalline Films Grown onto {YAlO$_3$} Substrates: Raman Spectroscopy Study toward Visualization of Mechanical Stress},
journal = {J. Phys. Chem. C},
volume = {125},
number = {29},
pages = {16279-16288},
year = {2021},
doi = {10.1021/acs.jpcc.1c03922},
URL = {https://doi.org/10.1021/acs.jpcc.1c03922},
}

@article{Ambrosio-2000,
title = {Crystalline {YAlO$_3$} as a novel window for photodetectors},
journal = {Nucl. Instr. and Meth. A},
volume = {454},
number = {1},
pages = {221-226},
year = {2000},
doi = {10.1016/S0168-9002(00)00831-7},
url = {https://www.sciencedirect.com/science/article/pii/S0168900200008317},
author = {C. {D'Ambrosio} and F. {De Notaristefani} and H. Leutz and D. Puertolas and E. Rosso},
}

@article{Uchiyama-2003,
    author = {Uchiyama, H. and Aizawa, H. and Katsumata, T. and Komuro, S. and Morikawa, T. and Toba, E.},
    title = {Fiber-optic thermometer using {Cr-doped} {YAlO$_3$} sensor head},
    journal = {Rev. Sci. Instrum.},
    volume = {74},
    number = {8},
    pages = {3883-3885},
    year = {2003},
    month = {08},
    issn = {0034-6748},
    doi = {10.1063/1.1589582},
    url = {https://doi.org/10.1063/1.1589582},
}

@article{Noginov-1998,
author = {M. A. Noginov and N. Noginova and M. Curley and N. Kukhtarev and H. J. Caulfield and P. Venkateswarlu and G. B. Loutts},
journal = {J. Opt. Soc. Am. B},
number = {5},
pages = {1463--1468},
publisher = {Optica Publishing Group},
title = {Optical characterization of {Mn:YAlO$_3$}: {Material} for holographic recording and data storage},
volume = {15},
month = {May},
year = {1998},
url = {https://opg.optica.org/josab/abstract.cfm?URI=josab-15-5-1463},
doi = {10.1364/JOSAB.15.001463},
}

@article{Loutts-1998,
  title = {Manganese-doped yttrium orthoaluminate: A potential material for holographic recording and data storage},
  author = {Loutts, G. B. and Warren, M. and Taylor, L. and Rakhimov, R. R. and Ries, H. R. and Miller, G. and Noginov, M. A. and Curley, M. and Noginova, N. and Kukhtarev, N. and Caulfield, H. J. and Venkateswarlu, P.},
  journal = {Phys. Rev. B},
  volume = {57},
  issue = {7},
  pages = {3706--3709},
  numpages = {0},
  year = {1998},
  month = {Feb},
  publisher = {American Physical Society},
  doi = {10.1103/PhysRevB.57.3706},
  url = {https://link.aps.org/doi/10.1103/PhysRevB.57.3706}
}

@article{Yamamoto-2022,
title = {Development of a phoswich detector composed of {ZnS(Ag) and YAP(Ce)} for astatine-211 imaging},
journal = {Radiat. Meas.},
volume = {153},
pages = {106734},
year = {2022},
issn = {1350-4487},
doi = {10.1016/j.radmeas.2022.106734},
url = {https://www.sciencedirect.com/science/article/pii/S1350448722000300},
author = {Seiichi Yamamoto and Naoyuki Ukon and Kohshin Washiyama and Koki Hasegawa and Kei Kamada and Masao Yoshino and Akira Yoshikawa},
}

@article{Vittori-1999,
title = {Crystals and light collection in nuclear medicine},
journal = {Nucl. Phys. B},
volume = {78},
number = {1},
pages = {616-621},
year = {1999},
doi = {10.1016/S0920-5632(99)00614-3},
url = {https://www.sciencedirect.com/science/article/pii/S0920563299006143},
author = {F Vittori and F {de Notaristefani} and T Malatesta},
}

@article{Costa-1995,
title = {Design of a scattering polarimeter for hard {X-ray} astronomy},
journal = {Nucl. Instrum. Meth. A},
volume = {366},
number = {1},
pages = {161-172},
year = {1995},
issn = {0168-9002},
doi = {10.1016/0168-9002(95)00460-2},
url = {https://www.sciencedirect.com/science/article/pii/0168900295004602},
author = {E. Costa and M. N. Cinti and M. Feroci and G. Matt and M. Rapisarda},
}

@article{Song-2023,
title = {High efficiency cryogenic {CW Nd:YAlO3} laser with dual wavelengths at 1072 and 1079 nm},
journal = {Opt. Laser Technol.},
volume = {163},
pages = {109397},
year = {2023},
issn = {0030-3992},
doi = {10.1016/j.optlastec.2023.109397},
url = {https://www.sciencedirect.com/science/article/pii/S0030399223002906},
author = {Yan-Jie Song and Nan Zong and Zhi-Min Wang and Zhong-Zheng Chen and Shen-Jin Zhang and Xiao-Jun Wang and Yong Bo and Qin-Jun Peng},
}

@article{Kishimoto-2003,
title = {Properties of a {YAP:Ce detector for high-energy X-ray} counting experiments},
journal = {Nucl. Instrum. Meth. A},
volume = {508},
number = {3},
pages = {425-433},
year = {2003},
issn = {0168-9002},
doi = {10.1016/S0168-9002(03)01655-3},
url = {https://www.sciencedirect.com/science/article/pii/S0168900203016553},
author = {Shunji Kishimoto and Tatsuo Yamamoto},
}

@article{Aggarwal-2005,
    author = {Aggarwal, R. L. and Ripin, D. J. and Ochoa, J. R. and Fan, T. Y.},
    title = {Measurement of thermo-optic properties of {Y$_3$Al$_5$O$_{12}$, Lu$_3$Al$_5$O$_{12}$, YAIO$_3$, LiYF$_4$, LiLuF$_4$, BaY$_2$F$_8$, KGd(WO$_4$)$_2$, and KY(WO$_4$)$_2$} laser crystals in the {$80-300$ K} temperature range},
    journal = {J. Appl. Phys.},
    volume = {98},
    number = {10},
    pages = {103514},
    year = {2005},
    month = {11},
    issn = {0021-8979},
    doi = {10.1063/1.2128696},
    url = {https://doi.org/10.1063/1.2128696},
}

@article{Petit-2010,
    author = {Petit, Johan and Viana, Bruno and Goldner, Philippe and Roger, Jean-Paul and Fournier, Danièle},
    title = {Thermomechanical properties of {Yb$^{3+}$} doped laser crystals: Experiments and modeling},
    journal = {J. Appl. Phys.},
    volume = {108},
    number = {12},
    pages = {123108},
    year = {2010},
    month = {12},
    issn = {0021-8979},
    doi = {10.1063/1.3520216},
    url = {https://doi.org/10.1063/1.3520216},
}

@article{Song-2020,
author = {Yanjie Song and Nan Zong and Ke Liu and Zhimin Wang and Xiaojun Wang and Yong Bo and Qinjun Peng and Zuyan Xu},
journal = {Opt. Mater. Express},
number = {7},
pages = {1522--1530},
publisher = {Optica Publishing Group},
title = {Temperature-dependent thermal and spectroscopic properties of {Yb:YAlO$_3$} perovskite crystal for a cryogenically cooled near {IR} laser},
volume = {10},
month = {Jul},
year = {2020},
url = {https://opg.optica.org/ome/abstract.cfm?URI=ome-10-7-1522},
doi = {10.1364/OME.392500},
}

@article{Xiang-2015,
title = {Theoretical investigations on mechanical anisotropy and intrinsic thermal conductivity of {YbAlO$_3$}},
journal = {J. Eur. Ceram. Soc.},
volume = {35},
number = {5},
pages = {1549-1557},
year = {2015},
issn = {0955-2219},
doi = {10.1016/j.jeurceramsoc.2014.12.008},
url = {https://www.sciencedirect.com/science/article/pii/S0955221914006657},
author = {Huimin Xiang and Zhihai Feng and Yanchun Zhou},
}

@article{morelli-1992,
doi = {10.1557/JMR.1992.2492},
url = {https://doi.org/10.1557/JMR.1992.2492},
year = {1992},
volume = {7},
number = {9},
pages = {2492},
author = {D. T. Morelli},
title = {Thermal conductivity of high temperature superconductor substrate materials: Lanthanum aluminate and neodymium aluminate},
journal = {J. Mater. Res.},
}

@article{buryy_thermal_2010,
doi = {10.1088/0953-8984/22/5/055902},
url = {https://dx.doi.org/10.1088/0953-8984/22/5/055902},
year = {2010},
month = {jan},
publisher = {},
volume = {22},
number = {5},
pages = {055902},
author = {O Buryy and Ya Zhydachevskii and L Vasylechko and D Sugak and N Martynyuk and S Ubizskii and K D Becker},
title = {Thermal changes of the crystal structure and the influence of thermo-chemical annealing on the optical properties of {YbAlO$_3$} crystals},
journal = {J. Phys.: Condens. Matter},
}

@article{Diehl-1975,
title = {Crystal structure refinement of {YAlO$_3$}, a promising laser material},
journal = {Mater. Res. Bull.},
volume = {10},
number = {2},
pages = {85-90},
year = {1975},
issn = {0025-5408},
doi = {10.1016/0025-5408(75)90125-7},
url = {https://www.sciencedirect.com/science/article/pii/0025540875901257},
author = {R. Diehl and G. Brandt},
}

@article{Wu-2019b,
  title = {Antiferromagnetic ordering and dipolar interactions of {YbAlO$_3$}},
  author = {Wu, L. S. and Nikitin, S. E. and Brando, M. and Vasylechko, L. and Ehlers, G. and Frontzek, M. and Savici, A. T. and Sala, G. and Christianson, A. D. and Lumsden, M. D. and Podlesnyak, A.},
  journal = {Phys. Rev. B},
  volume = {99},
  issue = {19},
  pages = {195117},
  numpages = {8},
  year = {2019},
  month = {May},
  publisher = {American Physical Society},
  doi = {10.1103/PhysRevB.99.195117},
  url = {https://link.aps.org/doi/10.1103/PhysRevB.99.195117}
}

@article{Zhang-2023,
title = {Thermal and fluorescence spectrum properties of {Dy:YAlO$_3$} crystal influenced by doping concentrations},
journal = {Opt. Mater.},
volume = {138},
pages = {113633},
year = {2023},
issn = {0925-3467},
doi = {10.1016/j.optmat.2023.113633},
url = {https://www.sciencedirect.com/science/article/pii/S0925346723002057},
author = {Cong Zhang and Yunru Chen and Shihui Ma and Honghua Fan and Yonggui Yu and Zhanggui Hu and Ning Ye and Jiyang Wang and Yicheng Wu},
}

@article{Hofmeister-2010,
    author = {Hofmeister, Anne M.},
    title = "{Thermal diffusivity of oxide perovskite compounds at elevated temperature}",
    journal = {J. Appl. Phys.},
    volume = {107},
    number = {10},
    pages = {103532},
    year = {2010},
    month = {05},
    issn = {0021-8979},
    doi = {10.1063/1.3371815},
    url = {https://doi.org/10.1063/1.3371815},
}

@article{zhan_theoretical_2012,
author = {Zhan, Xun and Li, Zhen and Liu, Bin and Wang, Jingyang and Zhou, Yanchun and Hu, Zijun},
title = {Theoretical Prediction of Elastic Stiffness and Minimum Lattice Thermal Conductivity of {Y$_3$Al$_5$O$_{12}$, YAlO$_3$ and Y$_4$Al$_2$O$_9$}},
journal = {J. Am. Ceram. Soc.},
volume = {95},
number = {4},
pages = {1429-1434},
doi = {10.1111/j.1551-2916.2012.05118.x},
url = {https://ceramics.onlinelibrary.wiley.com/doi/abs/10.1111/j.1551-2916.2012.05118.x},
year = {2012}
}

@article{Martelli-2021,
  title = {Thermal diffusivity and its lower bound in orthorhombic {SnSe}},
  author = {Martelli, Valentina and Abud, Fabio and Jim\'enez, Julio Larrea and Baggio-Saitovich, Elisa and Zhao, Li-Dong and Behnia, Kamran},
  journal = {Phys. Rev. B},
  volume = {104},
  issue = {3},
  pages = {035208},
  numpages = {7},
  year = {2021},
  month = {Jul},
  publisher = {American Physical Society},
  doi = {10.1103/PhysRevB.104.035208},
  url = {https://link.aps.org/doi/10.1103/PhysRevB.104.035208}
}

@article{Slack-1973,
title = {Nonmetallic crystals with high thermal conductivity},
journal = {J. Phys. Chem. Solids},
volume = {34},
number = {2},
pages = {321-335},
year = {1973},
issn = {0022-3697},
doi = {org/10.1016/0022-3697(73)90092-9},
url = {https://www.sciencedirect.com/science/article/pii/0022369773900929},
author = {G. A. Slack},
}

@article{Zhydachevskii-2006,
doi = {10.1088/0953-8984/18/49/028},
url = {https://dx.doi.org/10.1088/0953-8984/18/49/028},
year = {2006},
month = {nov},
publisher = {},
volume = {18},
number = {49},
pages = {11385},
author = {Ya Zhydachevskii and D Galanciak and S Kobyakov and M Berkowski and A Kami\'{n}ska and A Suchocki and Ya Zakharko and A Durygin},
title = {Photoluminescence studies of {Mn$^{4+}$ 
ions in YAlO$_3$} crystals at ambient and high pressure},
journal = {J. Phys.: Condens. Matter},
}

@article{Senyshyn-2013,
doi = { 10.12693/APhysPolA.124.329},
url = {https://dx.doi.org/10.12693/APhysPolA.124.329},
year = {2013},
volume = {124},
number = {2},
pages = {329},
author = {A Senyshyn AND L Vasylechko},
title = {Low Temperature Crystal Structure Behaviour of Complex
Yttrium Aluminium Oxides {YAlO$_3$ and Y$_3$Al$_5$O$_{12}$}},
journal = {Acta. Phys. Pol. A},
}

@article{Clarke-2003,
doi = {10.1016/S0257-8972(02)00593-5},
url = {https://dx.doi.org/10.1016/S0257-8972(02)00593-5},
year = {2013},
volume = {163-164},
pages = {67},
author = {D R Clarke},
title = {Materials selection guidelines for low thermal conductivity thermal barrier coatings},
journal = {Surf. Coat. Technol.},
}

@article{Reuss-1929,
author = {Reuss, A.},
title = {Berechnung der {Flie\ss grenze von Mischkristallen auf Grund der Plastizit\"atsbedingung f\"ur Einkristalle}},
journal = {Z. Angew. Math. Mech.},
volume = {9},
number = {1},
pages = {49-58},
doi = {10.1002/zamm.19290090104},
url = {https://onlinelibrary.wiley.com/doi/abs/10.1002/zamm.19290090104},
year = {1929}
}

@article{Hill-1952,
doi = {10.1088/0370-1298/65/5/307},
url = {https://dx.doi.org/10.1088/0370-1298/65/5/307},
year = {1952},
month = {may},
publisher = {},
volume = {65},
number = {5},
pages = {349},
author = {R Hill},
title = {The Elastic Behaviour of a Crystalline Aggregate},
journal = {Proc. Phys. Soc. A},
}

@book{Voigt-1928,
author = {W Voigt},
title = {Lehrbuch der Kristallphysik},
publisher = {Teubner, Leipzig},
year = {1928},
}

@book{Berman-1976,
author = {Robert Berman},
title = {Thermal conduction in solids},
publisher = {Oxford University Press},
year = {1976},
}

@book{Vasylechko-2009,
title = {Handbook on the Physics and Chemistry of Rare Earths},
chapter = { Perovskite-Type Aluminates and Gallates},
volume = {39},
author = {L.~Vasylechko AND A.~Senyshyn AND U.~Bismayer},
editor = { K.~A.~{Gschneidner, Jr.} AND J.-C.~G.~B\"unzli AND V.~K.~Pecharsky},
publisher = {Elsevier, North-Holland},
pages = {113--295},
year = {2009},
}

@article{Callaway-1959,
	title = {Model for lattice thermal conductivity at low temperatures},
	volume = {113},
	url = {https://link.aps.org/doi/10.1103/PhysRev.113.1046},
	doi = {10.1103/PhysRev.113.1046},
	number = {4},
	journal = {Phys. Rev.},
	author = {Callaway, Joseph},
	month = feb,
	year = {1959},
	pages = {1046--1051},
}

@article{ McCurdy-1970,
Author = {McCurdy, A. K. and Maris, H. J. and Elbaum, C.},
Title = {Anisotropic heat conduction in cubic crystals in the boundary scattering
   regime},
Journal = {Phys. Rev. B},
Year = {1970},
Volume = {2},
Number = {10},
Pages = {4077-4083},
Month = {NOV 15},
DOI = {10.1103/PhysRevB.2.4077},
url = {https://dx.doi.org/10.1103/PhysRevB.2.4077},
}

@article{Schnelle-2001,
doi = {10.1088/0022-3727/34/6/302},
url = {https://dx.doi.org/10.1088/0022-3727/34/6/302},
year = {2001},
month = {mar},
publisher = {},
volume = {34},
number = {6},
pages = {846},
author = {W Schnelle and  R Fischer and  E Gmelin},
title = {Specific heat capacity and thermal conductivity of
{NdGaO$_3$ and LaAlO$_3$} single crystals at low temperatures},
journal = {J. Phys. D: Appl. Phys.},
}

@article{Slack-1962,
Author = {Slack, G. A.},
Title = {Thermal Conductivity of {MgO}, {Al$_2$O$_3$}, {MgAl$_2$O$_4$}, and {Fe$_3$O$_4$} crystals from 3 Degrees to 300 Degrees {K}},
Journal = {Phys. Rev.},
Year = {1962},
Volume = {126},
Number = {2},
Pages = {427-\&},
url = {https://dx.doi.org/10.1103/PhysRev.126.427},
DOI = {10.1103/PhysRev.126.427},
ISSN = {0031-899X},
Unique-ID = {WOS:A19621469C00070},
}

@article{Palmer-1997,
  title = {Thermal conductivity of germanium crystals with different isotopic compositions},
  author = {Asen-Palmer, M. and Bartkowski, K. and Gmelin, E. and Cardona, M. and Zhernov, A. P. and Inyushkin, A. V. and Taldenkov, A. and Ozhogin, V. I. and Itoh, K. M. and Haller, E. E.},
  journal = {Phys. Rev. B},
  volume = {56},
  issue = {15},
  pages = {9431--9447},
  numpages = {0},
  year = {1997},
  month = {Oct},
  publisher = {American Physical Society},
  doi = {10.1103/PhysRevB.56.9431},
  url = {https://link.aps.org/doi/10.1103/PhysRevB.56.9431}
}

@article{Holland-1963,
  title = {Analysis of Lattice Thermal Conductivity},
  author = {Holland, M. G.},
  journal = {Phys. Rev.},
  volume = {132},
  issue = {6},
  pages = {2461--2471},
  numpages = {0},
  year = {1963},
  month = {Dec},
  publisher = {American Physical Society},
  doi = {10.1103/PhysRev.132.2461},
  url = {https://link.aps.org/doi/10.1103/PhysRev.132.2461}
}

@article{Allen-2013,
  title = {Improved {Callaway} model for lattice thermal conductivity},
  author = {Allen, Philip B.},
  journal = {Phys. Rev. B},
  volume = {88},
  issue = {14},
  pages = {144302},
  numpages = {5},
  year = {2013},
  month = {Oct},
  publisher = {American Physical Society},
  doi = {10.1103/PhysRevB.88.144302},
  url = {https://link.aps.org/doi/10.1103/PhysRevB.88.144302}
}

@article{Mokhtari-2025,
  title = {1/5 and 1/3 magnetization plateaux in the spin 1/2 chain system {YbAlO$_3$}},
  author = {P. Mokhtari AND S. Galeski AND U. Stockert AND  S. E. Nikitin AND R. Wawrzy\'nczak AND R. K\"uchler AND M. Brando AND L. Vasylechko AND O. A. Starykh AND E. Hassinger},
  journal = {Phys. Rev. Lett.},
  volume = {135},
  issue = {7},
  pages = {076704},
  numpages = {6},
  year = {2025},
  month = {Aug},
  publisher = {American Physical Society},
  doi = {10.1103/grfl-37g2},
  url = {https://link.aps.org/doi/10.1103/grfl-37g2}
}

@misc{TOGO-1,
  title = {Ab-initio phonon calculation for {YAlO$_3$ / Pnma (62) / materials id 3792}},
  author = {Atsushi Togo},
  publisher = {NIMS},
  year = {},
  url = {https://mdr.nims.go.jp/datasets/7f019aa0-4896-4060-805a-168e881af6de},
}

@misc{data,
  author = {Mokhtari, P. AND Stockert, U. AND Nikitin, S. AND Vasylechko, L. AND Brando, M.
  AND Hassinger, E.},
  title = {Data sets {''Low-temperature thermal conductivity of the substrate material YAlO3 and its unconventional sister compound YbAlO3'', Technische Universit\"at Dresden, 2025, https://doi.org/10.25532/OPARA-949}},
  url = {https://doi.org/10.25532/OPARA-949},
}

\clearpage

\end{document}